
%
\input harvmac

\def\Titleh#1#2{\nopagenumbers\abstractfont\hsize=\hstitle\rightline{#1}%
\vskip .5in\centerline{\titlefont #2}\abstractfont\vskip .5in\pageno=0}
%
\def\UAa{Department of Physics and Astronomy}
\def\UAb{The University of Alabama}
\def\UAc{\it Box 870324, Tuscaloosa, AL 35487-0324, USA}
\def\UAIa{Department of Physics}
\def\UAIb{Texas A\&I University}
\def\UAIc{\it Kingsville, TX 78363, USA}
%
%
\def\hb{\hfil\break}

\catcode`\@=11 

\def\lsim{\mathrel{\mathpalette\@versim<}}
\def\gsim{\mathrel{\mathpalette\@versim>}}
\def\@versim#1#2{\vcenter{\offinterlineskip
    \ialign{$\m@th#1\hfil##\hfil$\crcr#2\crcr\sim\crcr } }}
\def\boxit#1{\vbox{\hrule\hbox{\vrule\kern3pt
      \vbox{\kern3pt#1\kern3pt}\kern3pt\vrule}\hrule}}

\def\cl{\centerline}
\def\etal{{\it et al.}}

\def\t1{{\tilde 1}}

\def\tg{{\tilde g}}

\def\a3{{\alpha_3}}

\def\GeV{\,{\rm GeV}}

\def\NPB#1#2#3{Nucl. Phys. B {\bf#1} (19#2) #3}
\def\PLB#1#2#3{Phys. Lett. B {\bf#1} (19#2) #3}

\def\PRD#1#2#3{Phys. Rev. D {\bf#1} (19#2) #3}

\def\UAHEP#1{Alabama preprint UAHEP#1}
%
%
\nref\EBP{For recent review see\hfil\break
J. Ellis, CERN preprint CERN-TH.6193/91, (to appear in the Proceedings
of the LP-HEP 91 Conference, Geneva, 1991);\hb
S. Bethke and J.E. Pilcher, Ann. Rev. Nucl. Part. Sci. {\bf 42}, 1992.}
\nref\GKL{S.G. Gorishny, A.L. Kataev and S.A. Larin, \PLB{259}{91}{144}.}
\nref\BNP{E. Braaten, S. Narison and A. Pich, \NPB{373}{92}{581};\hb
A. Pich, in Heavy Flavors, eds. A.J. Buras and M. Lindner, World
Scientific, 1991.}
\nref\DataG{Particle Data Group, Phys. Rev. D {\bf 45}, No. 11 (1992).}
\nref\BES{BES Collaboration, talk presented at the APS meeting,
April 22, 1992.}
\nref\HEB{T. Hebbeker, Aachen preprint PITHA 91/17, (to appear in
the Proceedings of the LP-HEP 91 Conference, Geneva, 1991).}
\nref\MAR{W. Marciano, \PRD{45}{92}{721}.}
\nref\ALE{ALEPH Collaboration, D. Decamp \etal, \PLB{279}{92}{411}.}
\nref\CLE{CLEO Collaboration, R. Ammar \etal, \PRD{45}{92}{3976}.}
\nref\LOU{L. Clavelli, \UAHEP{921}, (Phys. Rev. D in press).}
\nref\CCY{L. Clavelli, P.W. Coulter, and Kajia Yuan, \UAHEP{924}.}
\nref\TT{T.N. Truong, preprint KEK-TH-330, 1992.}
\leftline{\titlefont THE UNIVERSITY OF ALABAMA}
\Titleh{\vbox{\baselineskip12pt\hbox{UAHEP9211}}}
{\vbox{\cl{Supersymmetry and Tau Decay}}}
\cl{L. CLAVELLI$^{(a)}$, P.H. COX$^{(b)}$, and KAJIA YUAN$^{(a)}$}
\bigskip
\cl{$^{(a)}$\UAa}
\cl{\UAb}
\cl{\UAc}
\bigskip
\cl{$^{(b)}$\UAIa}
\cl{\UAIb}
\cl{\UAIc}
\bigskip
\bigskip
\bigskip
\cl{ABSTRACT}
\bigskip
The impact of the new tau decay data on the various $\tau$ puzzles
and on the possibility of approximate supersymmetry is discussed.
The most economical solution of the problems in $\tau$ decay
and that favored by recent new data supports the existence of
gluinos below one GeV.
\bigskip
\Date{July, 1992}

Accurate data \EBP\ on the interactions of the tau lepton have
contributed to the impressive confirmation of the standard model but
have at the same time created several apparent contradictions with
important implications for QCD and grand unification.
As an example of the former, the branching ratio of the $Z^0$ into
$\tau^{+}\tau^{-}$ relative to $e^{+}e^{-}$ is a test of lepton
universality in the neutral current coupling at the one percent level
and that of the $W$ into $\tau{\bar \nu}_\tau$ relative to
$e{\bar \nu}_e$ confirms universality in the charged current
coupling at the few percent level. Hence to at least this accuracy
we must have
\eqn\I{\Gamma(\tau\rightarrow e{\nu_\tau}{\bar \nu}_e)=
\Gamma(\mu\rightarrow e{\nu_\mu}{\bar \nu}_e)
\Bigl({m_\tau\over m_\mu}\Bigr)^5,}
or
\eqn\II{B_\tau(e)={\tau_\tau\over \tau_\mu}
\Bigl({m_\tau\over m_\mu}\Bigr)^5,}
relating the electronic branching ratio of the $\tau$ to the masses and
lifetimes of the $\tau$ and $\mu$. $B_\tau(e)$ provides a potentially
important measure of the strong coupling constant in the low energy
regime since the standard model with standard particle content
predicts \GKL\
\eqn\III{R_\tau=B_\tau(e)^{-1}-1.973=
3.0582\Bigl[0.994+{\a3\over \pi}+5.2023\Bigl({\a3\over \pi}\Bigr)^2
+26.366\Bigl({\a3\over \pi}\Bigr)^3+\cdots\Bigr].}
Non-perturbative contributions included in the bracket in Eq. \III\ have
been estimated \BNP\ to be less than 1\%, making $\tau$ decay a
potentially reliable low energy measure of the strong coupling constant.
Because of the convergence of the renormalization group family of curves
a reliable low energy measure pins down the strong coupling constant
more effectively than a high energy measure of the same accuracy and
reliability.

This leads us to the first problem in the tau data, namely that the data
does not agree with Eq. \II. Since lepton universality is well confirmed
to the required accuracy, the experimental value of either the electronic
branching ratio or the $\tau$ lifetime must be in error by three standard
deviations assuming the other is correctly measured. That is, under this
assumption and using the Particle Data Group averages \DataG, either
$\tau_\tau$ is incorrectly measured and
\eqn\IV{B_\tau(e)=0.1785\pm 0.0029\quad
R_\tau=3.63\pm 0.09\quad  \a3(m_\tau)=0.33\pm 0.03}
{\it or} $B_\tau(e)$ is incorrectly measured and
\eqn\V{\tau_\tau=(3.05\pm 0.06)\times 10^{-13} {\rm sec}\quad
R_\tau=3.28\pm 0.10\quad  \a3(m_\tau)=0.181\pm 0.057}
The new Beijing measurement \BES\ of the $\tau$ mass, which is several
standard deviations lower than the world average, would increase the
$R_\tau$ of Eq. \V\ by 3\%, still not enough to eliminate the
discrepancy. The strong coupling constant of Eq. \IV, when extrapolated
to the $Z^0$ mass scale assuming standard particle content,
agrees with the world average value \HEB\
\eqn\VI{\a3(M_Z) =0.113\pm 0.003}
whereas Eq. \V\ would lead under the same assumption to a value
disagreeing with Eq. \VI\ and with the idea of a minimal SUSY grand
unification with all SUSY partners in the 100 GeV to 1 TeV region.
For these reasons, the prevalent attitude towards the discrepancy has
been that the lifetime measurement is for some reason erroneous \MAR;
an alternative approach is suggested below.

A second problem in $\tau$ decay is the so-called ``1-prong problem",
namely the world average data does not respect the identity (again
assuming lepton universality)
\eqn\VII{B_\tau(1\ {\rm prong})=1.973B_\tau(e)+
B_\tau({\rm hadrons}, 1\ {\rm prong}).}
The data \DataG\ has, in addition to the electronic branching ratio of
Eq. \IV, total one-prong and hadronic one-prong branching ratios
\eqn\VIII{B_\tau(1\ {\rm prong})=0.859\pm 0.002,}
\eqn\VIV{B_\tau({\rm hadrons},1\ {\rm prong})=0.467\pm 0.011.}
Again, in order to satisfy Eq. \VII, one of the three experimental
quantities in that equation must be in error by three standard
deviations assuming the others to be correct.

One solution to the two $\tau$ puzzles would be for several of the
experimental quantities to be in error by one to two standard deviations
each in the appropriate direction to resolve the discrepancies. This is
in fact the solution suggested by recent ALEPH data \refs{\DataG,\ALE}
which has the lifetime two standard deviations below the world average
with the electronic and various one-prong hadronic branching ratios one
standard deviation above the world averages. This data has of course
somewhat larger errors than the world average values.
If however the ALEPH solution survives as the errors decrease then the
resulting $\a3(M_Z)$ will be somewhat lower than the world average value
of Eq. \VI. In addition there may be some new theoretical discrepancies
with various one-prong hadronic decays such as the
$\tau\rightarrow \nu_\tau + \pi$ or $K$ which are very reliably
predictable.

As an alternative one may note that a unique way to resolve the two
$\tau$ decay puzzles by revising a single experimental quantity is to
find the electronic branching ratio several standard deviations higher
than the world average. This most economical solution is in fact
supported by the latest CLEO publication \CLE\ which finds
\eqn\X{B_\tau(e)=0.192\pm 0.004.}
This presently represents the most accurate electronic branching ratio
measurement. It leads via Eq. \III\ to
\eqn\XI{R_\tau=3.24\pm 0.11\qquad \a3(m_\tau)=0.156\matrix {+0.059 \cr
-0.076 \cr}}
in good agreement with the corresponding values from the $\tau$ lifetime
measurement (Eq. \V). It also simultaneously resolves the one-prong
problem assuming again lepton universality.

Eventually of course the measurements of the various experimental groups
working on $\tau$ decay will have to converge. If the final consensus is
close to the solution of the ALEPH collaboration there will be some
discrepancy between the $\tau$ result and the world average values
of $\a3(M_Z)$. In addition the strong coupling constant from
$\tau$ decay will also be in disagreement with the values from
charmonium requiring an appeal to large non-perturbative
corrections to QCD in $J/\psi$ decay.

The main purpose of this paper is to point out that the CLEO result not
only resolves the two $\tau$ decay puzzles but is also in agreement with
quarkonia measurements of $\a3$ and, if the gluino mass is below 4 Gev
as suggested in Ref. \LOU, with SUSY unification and the world average
value of $\a3(M_Z)$.

Since the CLEO measurement of $R_\tau$ is consistent with the value from
the lifetime measurement we may average the two results to reduce the
error. The combined result is
\eqn\XII{R_\tau=3.26\pm 0.07\qquad \a3(m_\tau)=0.17\pm 0.04.}
In table 1 we compare this result for the strong coupling constant with
those from 3 charmonium measurements \LOU\ at nearby scales (chosen to
minimize known higher order corrections). The expected difference between
the various values of $\a3$ due to the running of the coupling constant
between various scales involved is negligible compared to the individual
experimental errors. If the CLEO result is confirmed we will have a
remarkable agreement between three physically different measures of the
strong coupling constant in the region below 2 GeV.
The $\a3$ values in table 1 are obtained assuming standard model particle
content in the low energy region. As emphasized in Ref. \LOU\ the
charmonium data on $\a3$ taken at face value is inconsistent with world
average values of this coupling at the $Z^0$ and is inconsistent with
minimal SUSY unification with a common SUSY threshold below 10 TeV.
One way to make the charmonium data, and therefore the new CLEO data
on $B_\tau(e)$, consistent with these higher energy considerations is
to have the gluino octet in or below the quarkonium region \LOU.
The consistency of such light gluinos with other unsucessful
experimental searches is discussed in Refs. \LOU\ and \CCY.

In Ref. \CCY, a fit to all the three-gluon decays of the vector
quarkonia was done as a function of the strong coupling constant and the
gluino mass. Gluino masses near zero or near 0.3 GeV gave acceptable
$\chi^2$ with the best-fit values of $\a3$ agreeing with Eq. \VI\ above
in both cases. Such light gluinos would be expected to contribute to the
hadronic $\tau$ decays modifying the coefficients of the
$\alpha^2_3$ terms and higher in Eq. \III\ due both to
virtual processes and to direct decays such as
\eqn\XIII{\tau\rightarrow \nu_\tau\ q\ {\bar q}\ \tg\ \tg\ .}
If the gluino mass is below $m_\tau/2$ some of the hadrons in
$\tau$ decay would then be expected to contain gluino components.
If both gluinos hadronize into the same hadron it may be impossible
to directly distinguish the final state of Eq. \XIII\
from conventional hadronic final states. If however the gluinos
hadronize into separate hadrons these will decay by photino emission
(assuming $R$ parity conservation) leading to missing energy in the
hadronic final state. Unfortunately it will be difficult to distinguish
this from the missing energy carried off by the neutrino.

It is relatively easy to correct Eq. \III\ to take into account the
lowest order contribution from gluinos. The coefficient of the quadratic
term in the bracket of Eq. \III\ has the following dependence on the
number of quark flavors below the $\tau$ \TT\
\eqn\XIV{C_2=6.3399-0.3792 n_f.}
Since the gluinos behave like quarks with a relative color and statistics
factor of 3, the contribution of gluinos to order $\alpha^2_3$ is
obtained by making in Eq. \XIV\ the replacement
\eqn\XV{n_f\rightarrow n_f+3n_{\tg}}
where $n_{\tg}$ is the effective number of gluino octets. In the minimal
SUSY model $n_{\tg}$ is unity for massless gluinos and falls to zero for
increasingly massive gluinos. Thus in lowest non-trivial order, the
effect of light gluinos is to raise the value of $\a3$ required to fit
the $\tau$ data by some 4.6\%. In the vector quarkonia, a light gluino
has the opposite effect of lowering the required value of $\a3$ \CCY.
If the gluinos are as massive as 0.5 GeV the phase space suppression
is liable to make this contribution negligible in $\tau$ decay as well
as in charmonium decay. Therefore, if the light gluino hypothesis is the
correct interpretation of the data, and if additional data confirms the
CLEO result and reduces the error, $\tau$ decay is likely to become a
very sensitive measure of the gluino mass. At present, of course, the
widespread experimental disagreement in $\tau$ decay parameters and the
consequent large errors do not justify extending our analysis to include
the effect of gluinos on the cubic term in Eq. \III. In addition, if the
experiments ultimately converge on a value for $B_\tau(e)$ near the
current world average, then $\tau$ decay would strongly disfavor light
gluinos.

\bigskip
\bigskip
\cl{\bf Acknowledgments}\nobreak
The authors would like to acknowledge several useful discussions on
this topic with Phil Coulter. This work has been supported in part by
the U.S. Department of Energy under Grant No. DE-FG05-84ER40141 and
by the Texas National Laboratory Research Commission under Grant
No. RCFY9155.

\bigskip
\bigskip
\bigskip
\input tables
\centerjust
\vbox{\tenpoint\noindent {\bf Table 1}: Comparison of the strong
coupling constant from $\tau$ decay with that of charmonium.}
\bigskip
\thicksize=1.0pt
\begintable
process |scale $\mu_S$|$\alpha_3(\mu_S)$ \cr
${\Gamma(\tau\rightarrow \nu_\tau + {\rm hadrons})\over
\Gamma(\tau\rightarrow \nu_\tau{\bar \nu}_ee)}$|$m_\tau=1.78\GeV$|
$0.17\pm 0.04$\cr
${\Gamma(J/\psi(1S)\rightarrow 3g)\over
\Gamma(J/\psi(1S)\rightarrow e^{+}e^{-})}$|$1.37\GeV$|
$0.191\pm 0.004$\cr
${\Gamma(J/\psi(2S)\rightarrow 3g)\over
\Gamma(J/\psi(2S)\rightarrow e^{+}e^{-})}$|$1.62\GeV$|
$0.226\pm 0.027$\cr
${\Gamma(\eta_c\rightarrow 2g)\over
\Gamma(J/\psi(1S)\rightarrow e^{+}e^{-})}$|$1.49\GeV$|
$0.170\pm 0.030$\endtable
\listrefs

\bye